**Title:**

# Closed-loop discovery of out-of-distribution processing protocols by evolutionary search and uncertainty-aware learning

**Authors:**

Yu Liu[1*], Stanislav Udovenko[2], Ching-Che Lin[3,4], Jaegyu Kim[4,5], Lane W. Martin[3,4,6], Susan Trolier-McKinstry[2] and Sergei V. Kalinin[1,7*]

[1] Department of Materials Science and Engineering, University of Tennessee, Knoxville, Tennessee 37996, USA

[2] Materials Science and Engineering Department, Materials Research Institute, the Pennsylvania State University, University Park, Pennsylvania 16802, USA

[3] Department of Materials Science and NanoEngineering, Rice University, Houston, TX 77005, USA

[4] Rice Advanced Materials Institute, Rice University, Houston, Texas 77005, USA

[5] Department of Materials Science and Engineering, University of California, Berkeley, Berkeley, California 94720, USA

[6] Departments of Chemistry and Physics and Astronomy, Rice University, Houston, Texas 77005, USA

[7] Physical Sciences Division, Pacific Northwest National Laboratory, Richland, Washington 99354, USA

* Corresponding author: yliu206@utk.edu, sergei2@utk.edu

## Abstract

Many materials and chemical systems exhibit history-dependent responses, where functional outcomes are governed not only by final-state variables but by the time-dependent sequence of fields, temperatures, stresses, or chemical potentials applied during operation. Discovering effective new protocols for materials optimization is therefore a high-dimensional discovery problem in the processing history spaces, where the control variable is an entire waveform or sample history, and conventional strategies to assess the same either remain confined to conservative interpolative families or become prohibitively measurement intensive. Here, ferroelectric thin films are used as a model system in which the driving protocol is the scanning-probe-microscopy-based tip-bias waveform and the outcome is the nonlinear electromechanical response. A closed-loop workflow is introduced that couples an evolutionary search (*i.e.*, mutation and crossover applied to a compact waveform representation) with uncertainty-aware deep kernel learning to generate, rank, and experimentally validate candidate protocols efficiently. Starting from a small set of experimentally accessible seed waveforms, the system discovers waveform families that enhance nonlinearity, essentially by de-aging the ferroelectric. Spatially resolved before/after measurements show that the best-performing waveforms produce only dispersed, small-amplitude perturbations of the existing domain structure, whereas the worst-performing waveforms drive a single large, contiguous switched region. This contrast indicates that strong nonlinear response is associated with selective activation of pre-existing, weakly pinned domain-wall segments rather than with the creation of new domains or with long-range irreversible switching. More broadly, this framework reframes protocol tuning as a route from optimization to discovery, enabling experimental generation of out-of-distribution processing histories that can be generalized to other high-dimensional control problems, including synthesis and annealing trajectories, battery formation/charging protocols, microstructure engineering, and dynamic discovery in molecular and chemical systems.

## I. Introduction

The functionality of materials is governed not only by composition and equilibrium structure, but also by history, *i.e.*, the processing and usage trajectory that brings a system to its operating state. This is the defining feature of problems ranging from thermal treatments in metallurgy[1,2] and glass formation[3] to electrochemical formation/charging protocols in batteries[4-6] and time-dependent field conditioning in functional oxides[7]. Historically, useful protocols were discovered either through detailed mechanistic understanding (when available) or through long, empirical campaigns. Purely data-driven optimization, while conceptually attractive, can be impractical because experiments are slow, expensive, noisy, and strongly state-dependent, so naïve exploration of protocol space quickly becomes prohibitive in terms of required experimental budgets. A further complication is that the nominal "reset" state of a sample is not always known or controllable. In ferroelectrics, for example, the aging state can depend on environmental conditions such as temperature fluctuations [8-10] that lie outside direct control. Nominally identical preparation procedures can therefore yield different starting conditions, adding to the state-dependence that any protocol-optimization campaign must address.

This setting imposes two requirements and a broader aim, none of which are fully addressed by standard materials-optimization workflows. First, the algorithm must propose candidates that map directly onto executable laboratory actions, such as bounded waveforms, annealing schedules, or pulse sequences, rather than onto idealized coordinates that require a separate translation step. Second, the algorithm must remain sample efficient when each trial modifies the material state and when the reward contains measurement noise, drift, and spatial heterogeneity. Third, beyond refining existing protocols, the algorithm should be able to generate genuinely new ones. Together these criteria motivate a closed-loop testbed in which candidate protocols can be generated, executed, scored, and used to retrain the model on a timescale short enough to evaluate the algorithm itself.

A common approach to high-dimensional optimization is to introduce a low-dimensional representation, for example by embedding process histories into a latent space learned by variational autoencoders (VAEs) and then performing a Bayesian or surrogate-guided search in that space[11-15]. This strategy is effective for interpolation among valid candidates, but it also tends to remain near the training distribution unless novelty is explicitly engineered[16,17]. In contrast, genetic algorithms naturally generate out-of-distribution candidates through mutation and recombination, and can be defined directly in the space of executable operations, making them conceptually well aligned with real experimental control[18-25]. Evolutionary search, however, is often computationally and measurement intensive, and can be brittle in the presence of experimental noise[26,27]. Here, a tractable alternative is operationalized by coupling evolutionary generation with uncertainty-aware learning, enabling broad exploration while maintaining sample efficiency.

This paradigm is demonstrated using scanning-probe microscopy as a localized processing platform, where the tip-bias waveform constitutes the processing protocol and the resulting change in ferroelectric nonlinear electromechanical response provides the experimental reward. A closed-loop workflow combines mutation/crossover applied to a compact waveform representation with deep-kernel learning to rank candidate protocols based on uncertainty in order to select the most informative experiments. Starting from a small set of experimentally accessible seed waveforms, the system discovers families of novel waveforms that enhance nonlinearity while largely preserving the mesoscale domain pattern, revealing that strong nonlinear response is governed by depinning domain walls from shallow potential wells rather than extensive irreversible switching. More broadly, this model system illustrates a route from optimization to discovery: the closed-loop generation and validation of processing pathways beyond the distribution of conventionally used protocols, a capability that is directly relevant to protocol discovery problems in synthesis, microstructure engineering, electrochemical cycling, and molecular and chemical systems.

## Background and prior work

A common strategy for high-dimensional optimization is to map to a low-dimensional latent space and perform Bayesian optimization (BO). For molecular design, this idea was popularized through continuous representations learned by variational autoencoders (VAEs) coupled to property predictors and latent-space search[28-31]. For process/protocol discovery, an analogous approach is to learn embedded trajectories (virtual process embedding) and then optimize in the corresponding latent space. The key advantage is that the optimizer operates in a lower-dimensional and often smoother space than the original representation. Because a VAE latent space is structured by reconstruction/regularization and the sampling distribution of the inputs, however, the induced objective landscape in latent space can remain complex and can lead to slow convergence. In molecular settings, this manifests as the latent space BO proposing invalid or non-synthesizable candidates when the optimizer moves far from the training distribution[13,16,20,26,27,32]; constrained or trust-region variants can be introduced specifically to mitigate this failure mode[26,27].

Deep-kernel learning (DKL) offers a complementary pathway in which the representation is learned jointly with the target by combining a neural feature extractor with a Gaussian process surrogate[27,29,30,33]. In this setting, the embedding is not static and is shaped by both the feature distribution and the observed objective values, which can yield faster convergence. This behavior has been demonstrated for process optimization benchmarks and extended to molecular discovery on the benchmark dataset in quantum chemistry[26] and molecular machine learning (QM9)[33], where DKL constructs "dynamic" embeddings aligned with molecular functionality rather than purely structural similarity. The principal limitation of standard DKL is that it is not generative, as it excels at ranking and selecting among candidates but does not, by itself, define a mechanism to propose fundamentally new candidates outside a prespecified pool.

One route to bridge this gap is to integrate reconstruction constraints with target-aware latent modeling so that the learned representation supports both generation and property-guided organization. Recent VAE–DKL-style formulations explicitly pursue this integration by coupling a VAE-latent representation with a DKL-style predictive structure[29]. Conceptually, these hybrids aim to retain the feasibility bias of a generative decoder while benefiting from the faster, target-aligned organization that comes from training representations against the objective. This hybrid model, however, also generalizes only within the distribution of the training data.

A second and operationally direct route to protocol discovery is to use evolutionary operators (*i.e.*, mutation/crossover) as a generative engine, while using an uncertainty-aware surrogate to make the search efficient[26,27,32,34]. Genetic algorithms (GAs) are attractive because their operators can be defined in the space of physically executable actions (*e.g.*, pulse primitives, sequence grammars, bounded-coefficient updates), which makes them naturally compatible with "sim2real" translation. Their weakness is efficiency: a naïve GA requires evaluating many candidates, which is prohibitive when evaluations are expensive and noisy. The GA–DKL framework addresses this by using GAs to generate broad candidate pools and DKL to rank candidates (*e.g.*, via an upper confidence bound, UCB) via data-efficient active learning and selecting a small batch for evaluation; thus retaining the out-of-distribution (OOD) generative capability of GAs while controlling experimental cost. Importantly, only a small subset of the GA-generated waveforms are used to train the DKL model, while predictive response and uncertainty are derived for the whole generation. In the present work, this logic is instantiated experimentally for waveform/protocol discovery in ferroelectrics by combining a VAE initialization manifold with a GA–DKL loop that proposes, ranks, and validates new tip-bias waveforms directly on the instrument.

## II. Experimental objective: ferroelectric nonlinearity

### a. Quantifying the nonlinear electromechanical response

Piezoelectric nonlinearity is the field-dependent departure of the electromechanical response from the small-signal linear regime. As the applied ac field increases, the measured piezoresponse no longer scales proportionally with drive amplitude because changes in the polarization state, motion of domain walls, and features of the local defect landscape begin to participate in the response[35-39]. This behavior is technologically important because the nonlinear susceptibility affects how ferroelectrics operate under realistic large-signal excitations, which in turn affects the functional performance of actuators, tunable capacitors, nonvolatile memories, and adaptive electronic devices. Microscopically, nonlinearity commonly originates from domain-wall bending and depinning, nucleation and growth of nanoscale switched regions, or phase-boundary motion[40-43], which in turn depend on defect- or charge-mediated internal fields, electrostatic and mechanical boundary conditions, and changes in local electromechanical coupling. Depending on the balance of these processes, the response can show field-dependent enhancement from domain-wall motion and local depinning, hysteretic and history-dependent domain growth, saturation at high drive, or irreversible switching and fatigue. Piezoelectric nonlinearity is a sensitive probe of how mesoscale domain structures and defects convert a time-dependent electrical stimulus into electromechanical functionality. In ferroelectrics, irreversible domain-wall motion supplies a substantial fraction of the usable dielectric and piezoelectric response. Aging progressively clamps this domain-wall population and thereby degrades the response over the device lifetime. The post-/pre-waveform nonlinearity ratio therefore serves as an operational measure of de-aging[8]. A waveform that maximizes this ratio electrically restores domain-wall mobility, and with it the extrinsic contribution to the response[44]. Notably, this restoration requires neither thermal annealing above the Curie temperature nor full repoling, neither of which is readily available in integrated devices.

Piezoelectric nonlinearity was examined in a $Pb_{0.995}(Zr_{0.45}Ti_{0.55})_{0.99}Nb_{0.01}O_3$ (PZT) thin-film model system using piezoresponse force microscopy (PFM); measurements were made between the tip and the bottom electrode (*e.g.*, on the bare piezoelectric surface without a top electrode). The sample structure, representative surface morphology, low-drive (0.1 V) piezoresponse contrast, and local switching response with a superimposed DC bias are summarized (Fig. 1a–d). Although the film was not macroscopically poled, as-grown PZT on $SrRuO_3/SrTiO_3$ develops a polydomain structure with finite local out-of-plane polarization (Fig. 1c). PFM therefore measures a local piezoelectric response within this native domain structure, and the nonlinearity metric defined below is referenced to this as-grown state rather than to a macroscopically poled one. These measurements establish the ferroelectric character of the film and provide the baseline domain structure on which waveform-induced changes are evaluated. The nonlinear electromechanical response was then quantified using two experimentally compatible readout schemes, both operated in the small-signal range of 0.1–1.2 V, which is well below the coercive voltage of 2.5–3 V (Fig. 1d) so that the response can be probed without driving macroscopic switching, with the linear baseline fitted between 0.1 and 0.36 V. In the first approach, the contact-resonance tuning curve was measured by sweeping the drive frequency at a series of drive voltages, and the peak piezoresponse amplitude at each drive voltage was obtained from a Lorentzian fit to the resonance (Fig. S1). In the second approach, the piezoresponse was monitored at a fixed offset of 10 kHz from resonance while dual-amplitude resonance tracking (DART) continuously followed the contact resonance during measurement. The DART-based readout is substantially faster, which makes it better suited for closed-loop optimization over many spatial locations.

The resonance-frequency shift (Fig. 1e and Fig. S4) indicates a reversible voltage-dependent change in the contact transfer function upon increasing and decreasing the AC drive amplitude ($V_{ac}$). Repeated voltage sweeps (Fig. S4) largely overlap, showing that this shift is reproducible over the measurement window rather than arising from progressive tip degradation or irreversible sample modification. Because fixed-frequency amplitude changes would therefore contain detuning artifacts, the raw amplitude$-V_{ac}$ curve cannot be directly interpreted as an absolute piezoelectric coefficient. Instead, DART tracking was used during the fast readout to define an operational, resonance-tracked effective nonlinear electromechanical response.

Across the 0.1–1.2 V small-signal range, this resonance-tracked response does not scale linearly with drive amplitude: above 0.6 V, it systematically exceeds the slope extrapolated from the lowest-drive points (Fig. 1e,f). The effective nonlinear response (ENL) was therefore defined as the deviation of the resonance-tracked piezoresponse-versus-drive curve from its low-drive linear trend, with the linear baseline fitted between 0.1 and 0.36 V. For each candidate waveform, ENL was measured before and after applying a 3.5 V maximum-amplitude bias waveform, using three 0.1–1.2 V sweeps for each readout. The optimization target was the post-/pre-waveform ENL ratio:

**Eq. (1).** ENL ratio = $ENL_{after}/ENL_{before}$

so that values larger than unity correspond to an enhancement of nonlinearity induced by the applied waveform. This ratio-based objective suppresses spatially varying baseline contrast and makes the optimization more robust to location-to-location differences in the starting state of the film. Both ENL readout protocols were tested during method development and were found to produce the same qualitative ranking of waveform performance. The faster DART-compatible readout was therefore adopted for the iterative searches. To minimize history effects from repeated switching, each ENL measurement was performed at a previously unmeasured location.

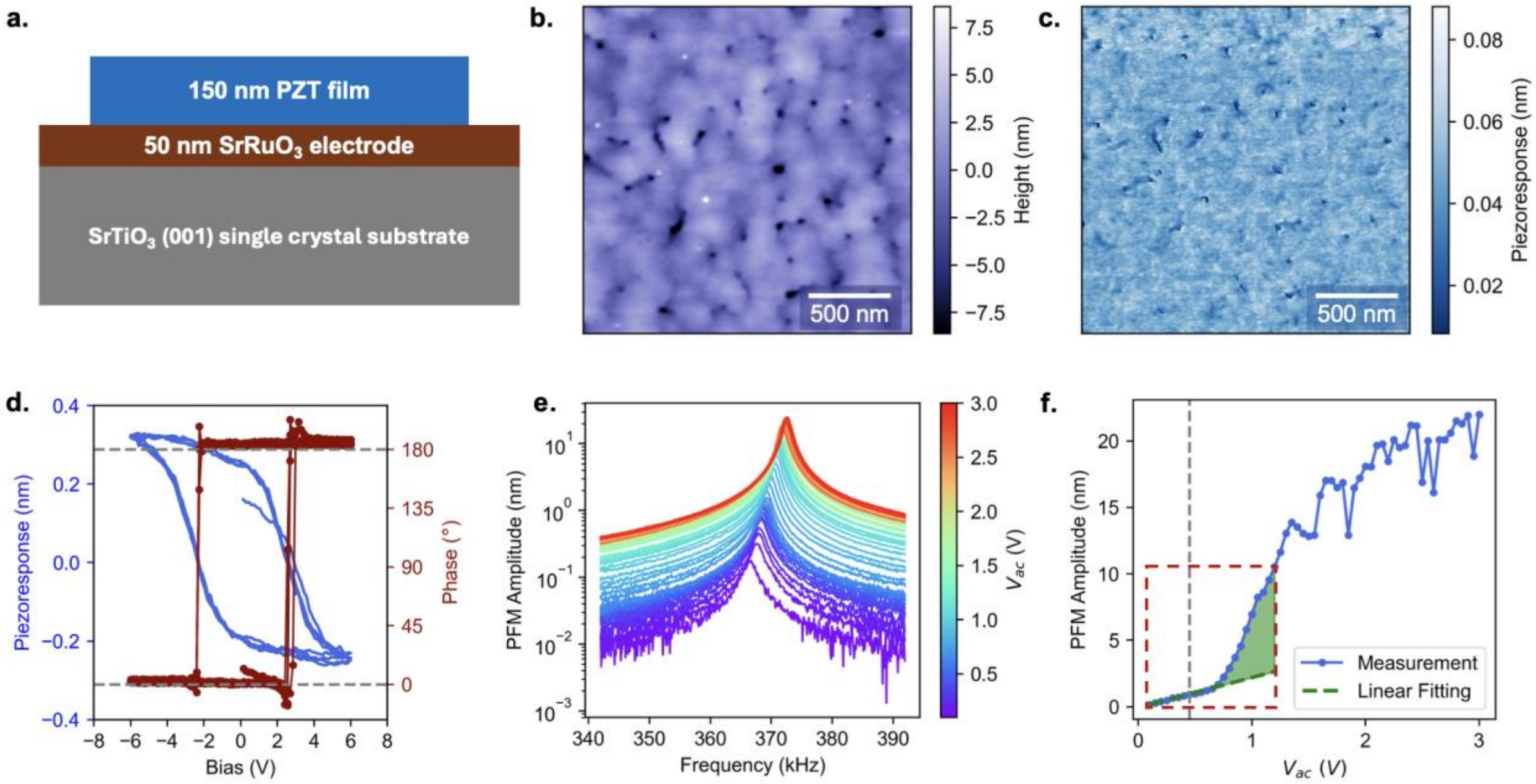

**Figure 1. Overview of the nonlinear response in ferroelectrics. a,** schematic structure of the $Pb_{0.995}(Zr_{0.45}Ti_{0.55})_{0.99}Nb_{0.01}O_3$ (PZT) thin-film sample. **b-c,** representative height and piezoresponse maps measured by DART-PFM at a drive voltage of 0.1 V. **d,** typical piezoresponse amplitude and phase hysteresis loops measured by DART-PFM switching spectroscopy. **e,** contact-resonance tuning curves acquired at different drive voltages from 0.1 V to 3 V. **f,** definition

of the nonlinear-response metric as the deviation of the drive-dependent piezoresponse from its low-drive linear baseline. Red dashed box indicates the small-signal measurement range (0.1 – 1.2 V) used to define ENL and the gray dashed line represent the cut-off drive for the linear baseline fitting (0.1 – 0.36 V).

### b. VAE to generate continuously varying candidates

Before launching the full GA-DKL optimization, a VAE was used to determine whether the experimentally measured ENL ratio varies smoothly across waveform shape space and whether interpolation between familiar waveforms can reveal useful new candidates. The VAE was initialized from 14 commonly used bias waveforms (Fig. 2a), each of which was tested experimentally 10 times at different locations to build the response statistics on aged, unpoled PZT films. These measurements indicate that waveform shape matters: the exponential growth and decay waveforms (No. 4-6, Fig. 2b) consistently produce larger ENL ratios than others; this is not unexpected, since these waveforms should be the most effective at depinning domain walls, and enhancing local nonlinearity.

To construct a continuous waveform manifold, each 4096-point waveform was standardized using a global mean and standard deviation and passed to a one-dimensional convolutional VAE. The encoder contained five stride-2 convolutional blocks with kernel size 9 and channels 16, 32, 64, 128, and 128, followed by linear layers that produced the mean and log-variance of a two-dimensional latent variable z. A mirrored transposed-convolution decoder reconstructed the waveform back to the original time domain. The model was trained by minimizing a reconstruction term together with a Kullback-Leibler regularization term[12,17],

**Eq. (2).** $$L_{VAE} = \left||x - \hat{x}|\right|^2 + \beta_{KL} \cdot D_{KL}[q(z|x)||N(0, I))]$$

where $x$ is the input waveform, $\hat{x}$ is its reconstruction by the decoder, $z$ is the two-dimensional latent variable, $q(z|x)$ is the Gaussian approximate posterior produced by the encoder (with mean and log-variance output heads), $N(0, I)$ is the standard normal prior on $z$ (with $I$ the identity covariance matrix), $D_{KL}[* \,||\, *]$ denotes the Kullback–Leibler divergence, and $\beta_{KL}$ is the weight that balances the reconstruction and regularization terms. Training used $\beta_{KL}$ = $10^{-3}$, AdamW optimization with a learning rate of $2 \times 10^{-4}$, a batch size of 8, and 3,000 training epochs. Because only 14 seed waveforms were available, the purpose of the VAE was not to learn an unrestricted generative distribution, but rather to provide a smooth, low-dimensional interpolation space that preserves experimentally realistic waveform shapes.

After training, the 2D latent space was sampled uniformly on a 21 × 21 grid, and each grid point was decoded into a new waveform candidate (Fig. 2c). Experimental testing of this VAE-generated library revealed a structured response landscape rather than isolated outliers (Fig. 2d): neighboring latent points often produced similar ENL ratios, while distinct regions of latent space yielded systematically stronger enhancement. This result shows that the waveform-response relationship is continuous enough (based on experience) to justify a model-based search, and it provides a physically reasonable set of starting candidates for the subsequent active-learning loop.

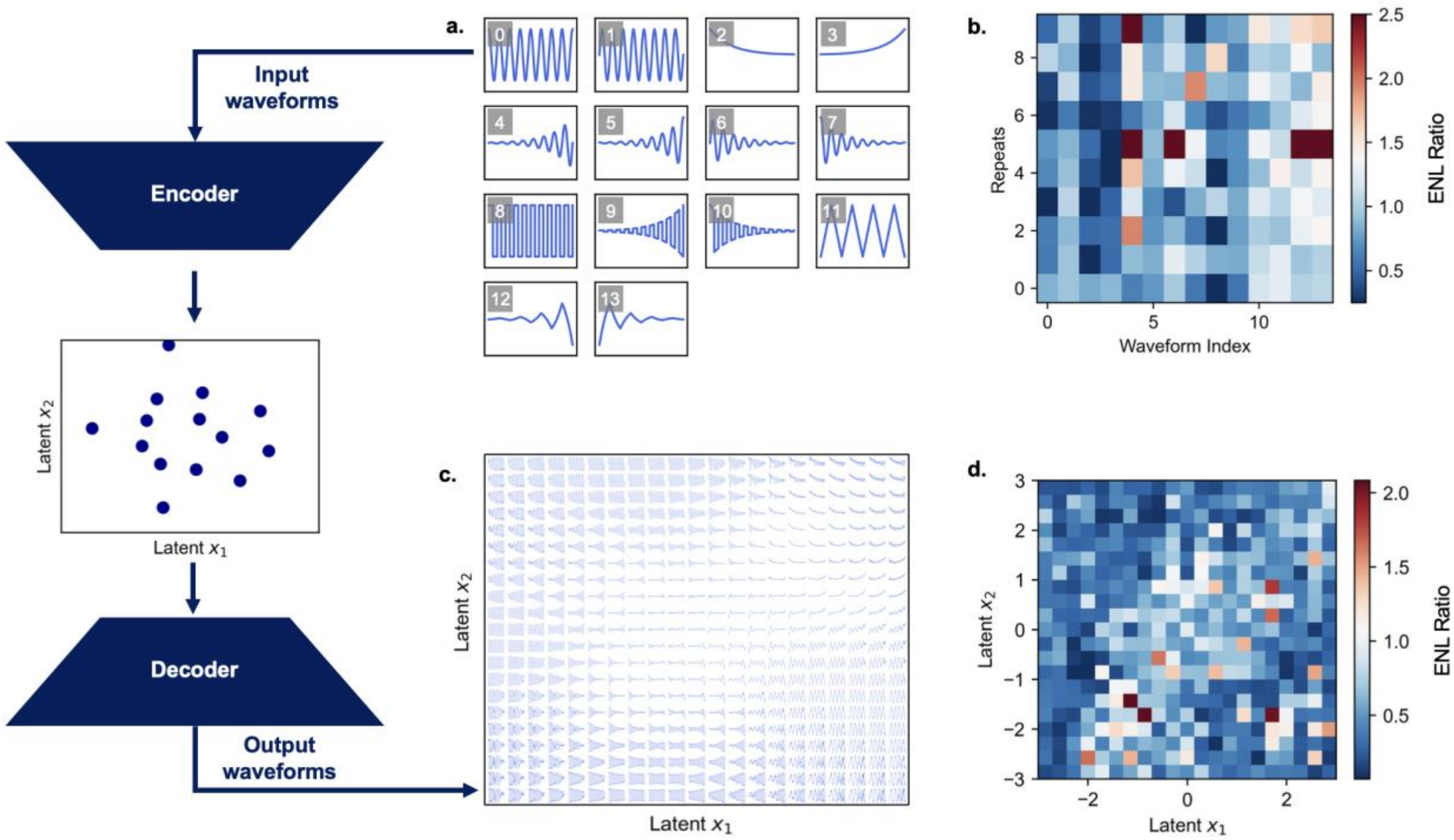

**Figure 2. Seeding ENL measurements generated by variational autoencoder (VAE). a,** fourteen bias waveforms used to initialize the VAE. **b,** experimentally measured ENL ratio before and after application of each seed waveform, with each waveform repeated at 10 new locations. **c,** a 2D VAE latent space built from the initial waveforms and sampled on a 21 × 21 grid to generate smoothly varying candidates. **d,** experimentally measured ENL ratios of the VAE-generated candidates, revealing structured regions of enhanced nonlinearity.

## III. Out-of-distribution waveform discovery by GA-DKL

The VAE library provides a structured starting manifold, but it does not by itself solve the much larger problem of discovering new waveform shapes that lie outside the distribution of the seed candidates, which is equivalent to optimizing over the infinite-dimensional space of all possible waveforms. This challenge is universal in molecular and materials discovery, where moving beyond the distribution defined by a small set of seed candidates is effectively equivalent to optimizing over the intractable space of all possible candidates.

Here, a closed-loop workflow was implemented combining a genetic algorithm for exploration with DKL for sample-efficient ranking (Fig. 3). The experimental campaign consisted of two seeding generations followed by 40 active-learning generations, with 32 experimentally tested waveforms in each generation. The 64 seed waveforms were randomly selected from the 21 × 21 VAE-generated candidate library. After seeding, every new generation was proposed by the GA-DKL loop and then verified directly in experiment.

For optimization, each candidate waveform $V(t)$ was mapped to a compact 33-parameter Fourier representation with one DC term and sine/cosine coefficients up to the 16$^{th}$ harmonic:

**Eq. (3).** $$V(t) = a_0 + \sum_{k=1}^{16}\left(a_k \cos\left(\frac{2\pi k}{T} t\right) + b_k \sin\left(\frac{2\pi k}{T} t\right)\right), \quad 0 \le t < T$$

where $T = 4$ s is the waveform duration. This representation preserves the global waveform shape while making genetic operations straightforward. After each experimental generation, the measured candidates were ranked by their ENL ratio and the highest-performing waveforms were used as parents. In the implementation used here, the search was restricted to the top-performing subset, namely the best eight experimentally measured waveforms in each generation. The corresponding GA was implemented using bounded coefficient-space operations, including a blend crossover[45]:

**Eq. (4).** $$c_i^{child} \sim U\left[c_{min,i} - \alpha \cdot d_i, c_{max,i} + \alpha \cdot d_i\right]$$

where $U[x, y]$ is the uniform sampling operator between $x$ and $y$, $c_i^{child}$ is the i$^{th}$ child Fourier coefficient, $c_i^{(1)}$ and $c_i^{(2)}$ are two parent Fourier coefficients, $c_{min,i} = min(c_i^{(1)}, c_i^{(2)})$, $c_{max,i} = max(c_i^{(1)}, c_i^{(2)})$, and $d_i = c_{max,i} - c_{min,i}$. The extrapolation parameter $\alpha = 0.5$ therefore allows offspring to lie up to 50% of the parental spread beyond either parent, which controls how aggressively the crossover explores outside the parent envelope. Mutations are introduced through additive Gaussian mutation:

**Eq. (5).** $$c_i^{mut} = c_i + \varepsilon_i, \ \varepsilon_i \sim N(0, \sigma^2)$$

where the mutation scale $\sigma = 0.15$ (in normalized coefficient units), and the probability of mutation is $p_{mut} = 0.7$. Parent waveforms were selected by tournament selection, and the best-performing candidates with the highest measured ENL were preserved through elitist retention. Before Fourier synthesis of the time-domain waveforms, all coefficients were clipped to their physically allowed bounds. Duplicate candidates were removed to maintain population diversity. Together, these operations generated a pool of 1,024 candidate waveforms for each active-learning cycle.

The DKL model was trained on all waveform-response pairs accumulated up to the current generation, rather than only the most recent batch. Waveforms were globally standardized and passed through a one-dimensional convolutional feature extractor consisting of four convolutional layers with channels 16, 32, 64, and 64, kernel size 9, stride 2, and Gaussian error linear unit (GELU)[46] activations, followed by adaptive average pooling and a two-layer head that mapped the waveform to a 32-dimensional latent descriptor. An exact Gaussian process with a constant mean and an automatic-relevance-determination radial basis function kernel was then trained on these latent descriptors using the exact marginal log-likelihood. In this work, this model was optimized for 200 iterations with Adam at a learning rate of $10^{-2}$.

To balance exploitation and exploration, the trained DKL surrogate ranked the 1,000 GA-generated candidates using an UCB acquisition function[26,34]:

**Eq. (6).** $UCB(x) = \mu(x) + \kappa \cdot \sigma(x), \kappa = 2$

where $\mu(x)$ and $\sigma(x)$ are the predicted mean and standard deviation of the ENL ratio. A greedy diversity filter was applied in the waveform space. Candidate waveforms were ranked by UCB score, and a candidate was included only if the Euclidean distance between its trace and every already selected trace was at least 5.0 in Fourier coefficient space. Waveforms with smaller down-sampled Euclidean distances were treated as near-duplicates and excluded from the same experimental batch. The top 32 ranked, diverse candidates were measured on the instrument, appended to the training archive, and used to retrain the DKL model for the next generation. This iterative loop links the generative breadth of the GA with the uncertainty-aware ranking of the DKL surrogate, allowing the experiment to explore broadly while still concentrating measurements on the most promising regions of waveform space.

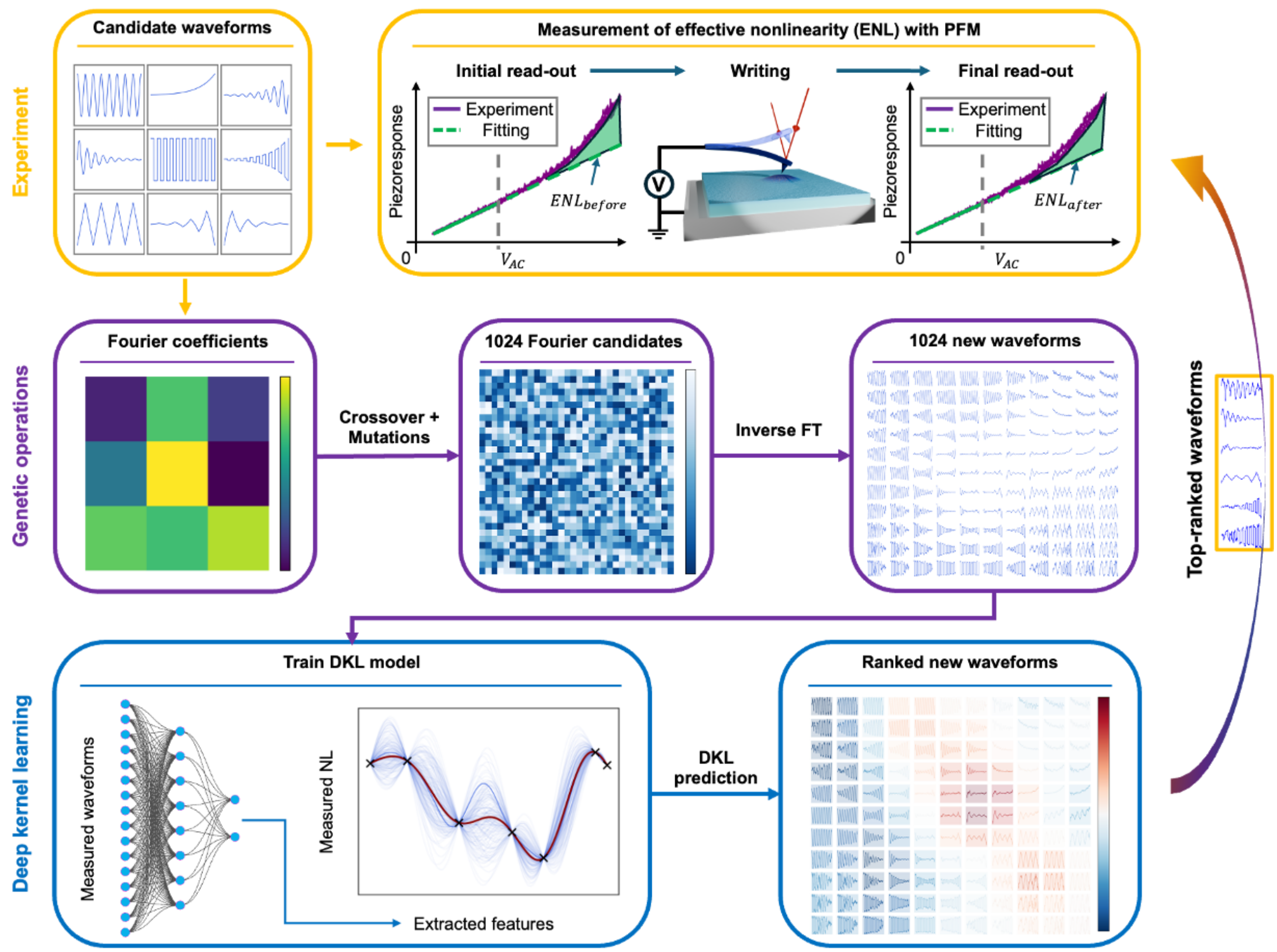

**Figure 3.** Detailed implementation of the GA-DKL workflow for waveform optimization. Candidate waveforms are first tested experimentally for their impact on ENL. The measured waveforms are parameterized by Fourier coefficients up to the 16$^{th}$ harmonic, enabling genetic operations such as crossover and mutation to generate a large pool of 1000 candidate waveforms. A DKL model trained on accumulated waveform–ENL pairs scores the candidate pool and selects the top 32 diverse candidates for experimental validation in the next generation. The newly measured responses are added back to the training set for iterative model refinement.

## IV. Interpretation of the GA-DKL results

The experimental results (Fig. 4) confirm that the closed-loop search progressively improves the response metric. The average ENL ratio increases over the course of the campaign (Fig. 4a), indicating that the optimizer is not merely finding rare high-value outliers but is shifting the full distribution of tested candidates toward stronger nonlinear behavior. At the same time, the best waveform identified in each generation evolves in a coherent manner rather than changing randomly from one generation to the next (Fig. 4b). This behavior is consistent with a smooth control landscape in which the algorithm learns a family of effective waveforms rather than a single isolated solution.

The best-performing waveforms point to a depinning-dominated origin of the enhanced ENL. Within a Rayleigh-type decomposition of the small-signal piezoresponse[36,37], the reversible part of domain-wall motion contributes primarily to the field-independent linear coefficient, while the irreversible part (*i.e.*, jumps of domain-wall segments out of shallow pinning potentials) is likely the dominant source of the field-dependent nonlinear contribution. Enhancing ENL therefore requires repeatedly accessing low-barrier depinning events of the existing weakly pinned domain-wall population, without driving the long-range coherent switching that would consume them. In contrast to a conventional triangular waveform, which applies a single monotonic bias ramp and is well known to favor growth of larger switched regions in ferroelectric thin films, the optimized waveforms contain asymmetric ramps, repeated local extrema, and multi-harmonic voltage modulations (Fig. 4c–g). These temporal features drive the local polarization through intermediate-field regimes multiple times within a single bias trace, an excitation pattern that is expected to favor small-amplitude depinning of the existing wall population while avoiding the long field exposure that would merge separate domains into larger switched regions. Selective activation of weakly pinned domain walls, rather than creation of new domain walls or long-range macroscopic switching, therefore provides the dominant physical route to the observed enhancement of ENL. From a practical standpoint, these results suggest waveform-engineered restoration as a maintenance operation for piezoelectric devices. In this scheme, a tailored, bounded-amplitude bias trace is applied in operando to recover the response lost to aging. As a result, a given strain or sensing output can be achieved at lower drive voltage. Moreover, the same search also identifies waveforms that suppress wall mobility through long-range switching. The workflow therefore provides bidirectional control over the extrinsic contribution: enhancement where maximum response is desired, and suppression where linearity and low loss are required.

The spatially resolved before/after measurements provide physical insight into what the optimizer is doing to the ferroelectric state. DART-PFM probes a tip-radius-limited, near-surface volume (approximately 30 nm laterally and 30–100 nm in depth at the drive levels used here), so the descriptors below refer to the projected near-surface domain population. Waveforms that yield the largest ENL ratio leave the mesoscale ferroelectric domain structure largely intact, with the change map (Fig. 4j) showing only dispersed, sign-balanced perturbations concentrated near the periphery of pre-existing domains. Within the lateral resolution of DART-PFM, this pattern does not strictly distinguish selective activation of pre-existing fine-scale features from partial fragmentation of larger domains near their boundaries; both routes are consistent with the data, and both preserve a high density of weakly pinned domain-wall segments available to contribute to the nonlinear response. By contrast, waveforms that give the smallest ENL ratio more often induce a single large, contiguous switched region in the change map (Fig. 4m), which is the signature of long-range coherent switching that consumes domain walls and leaves behind a region with reduced domain-wall length per unit area available for further nonlinear response. The qualitative contrast between Fig. 4h–j and Fig. 4k–m is therefore consistent with the Rayleigh-framework expectation that ENL is governed by the size of the irreversibly mobile domain-wall population accessible at small drive voltages.

The Fourier-coefficient evolution of the best-performing waveform in each generation further supports this interpretation (Fig. S2). Rather than converging to a trivial single-frequency drive, the optimized candidates retain contributions from several low- and intermediate-order harmonics, indicating that the useful control signal is encoded in a coordinated multi-harmonic waveform. In other words, the GA-DKL workflow is not simply finding a larger or sharper pulse; it is identifying a temporally structured drive that couples efficiently to small-amplitude, irreversible depinning of the existing weakly pinned domain-wall population. This result is consistent with prior studies of electrical de-aging in ferroelectrics, where time-dependent electric-field protocols can release aged or clamped domain-wall configurations[8]. Related work on pulsed and AC poling has also shown that non-DC field protocols can modify Rayleigh parameters associated with domain-wall mobility[44]. This is precisely the type of problem for which data-efficient active learning is valuable, because the underlying control variable is high dimensional, nonlinear, and experimentally expensive to evaluate. The same GA-DKL strategy was also validated on a $PbZr_{0.2}Ti_{0.8}O_3$/LSMO/$SrTiO_3$ (111) heterostructure, where the active-learning loop again improved the ENL ratio over successive generations (Fig. S3), supporting the transferability of the workflow beyond the primary PZT sample.

Taken together, the results indicate that waveform engineering can be used to bias ferroelectric systems toward a more nonlinear state without driving long-range irreversible switching, *i.e.*, toward a domain configuration whose existing weakly pinned domain-wall segments are repeatedly depinned rather than consumed. More broadly, the workflow demonstrates that experimental control waveforms can be treated as optimization variables in the same way that composition, temperature, or tip bias are treated in autonomous materials experiments. The same strategy should be extendable to other scanning-probe tasks in which the response depends sensitively on the temporal structure of the excitation, including domain writing, spectroscopy, and adaptive imaging protocols.

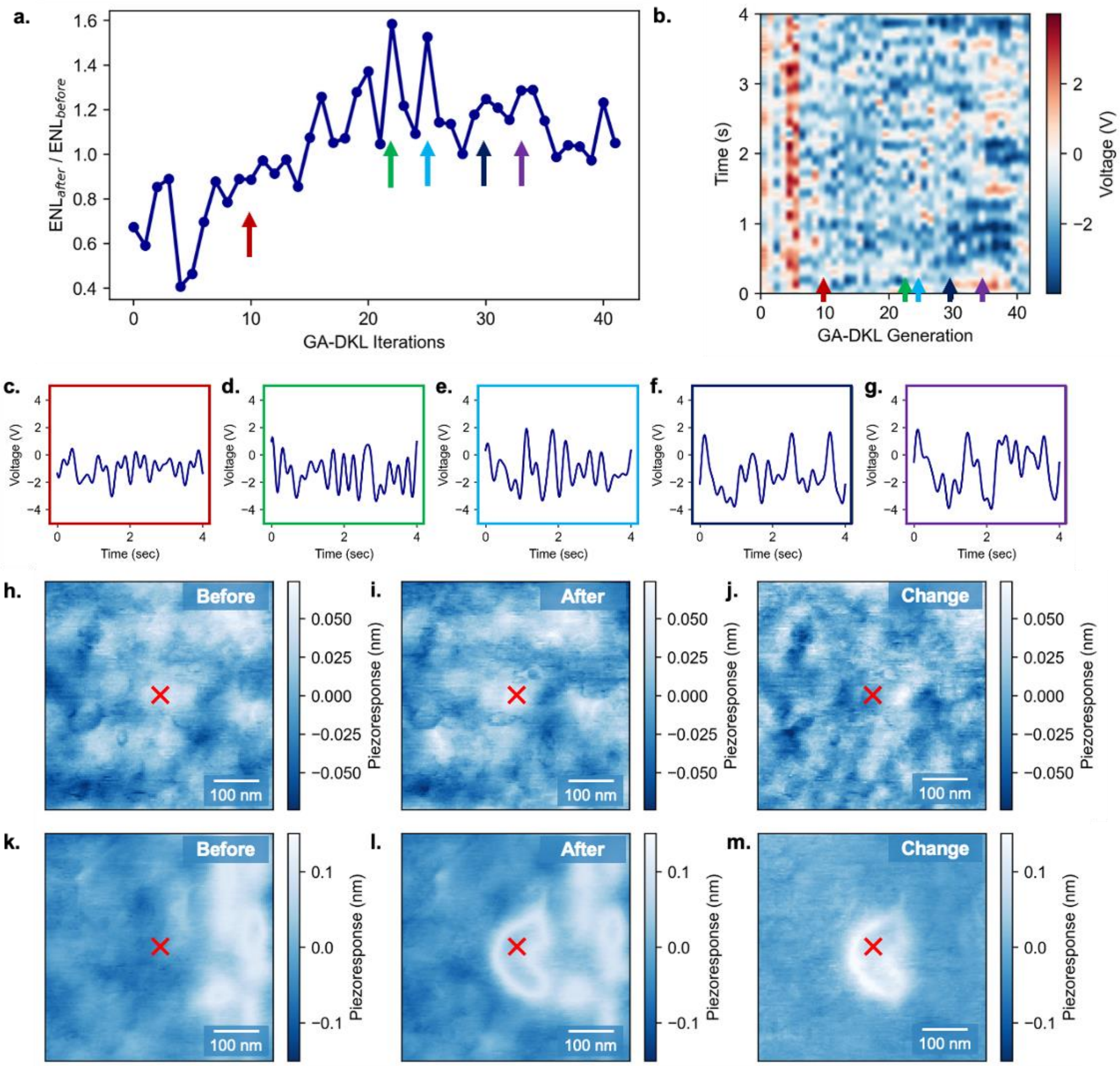

**Figure 4. Results of the GA-DKL experiment on $Pb_{0.995}(Zr_{0.45}Ti_{0.55})_{0.99}Nb_{0.01}O_3$ (PZT). a,** evolution of the averaged ENL ratio in each generation. The campaign contains two seeding generations and 40 active-learning generations, each with 32 waveform candidates. All the waveform candidates are shown **b,** evolution of the best waveform in each generation. **c–g,** best-performing voltage waveforms from selected generations highlighted in **a** and **b**, illustrating the evolution from early candidate waveforms toward temporally structured, multi-harmonic profiles discovered by the GA-DKL loop. **h-j,** example of a waveform that gives a large ENL ratio at generation 37. **h,** local piezoresponse map before waveform application; **i,** map after waveform application; **j,** the change map. The red cross marks the application location. **k-m,** analogous maps for a waveform that gives a small ENL ratio.

## V. Conclusion

The effective nonlinear response of a ferroelectric film can be optimized directly in experiment by treating the full-bias waveform as the control variable, to implement ferroelectric de-aging. This addresses a dual challenge that recurs across processing optimization: (i) establishing a fast experimental benchmarking system with repeatable feedback, and (ii) enabling tractable search in a high-dimensional protocol space. A VAE provides a compact and experimentally realistic initialization manifold, while the GA–DKL loop couples out-of-distribution candidate generation (via evolutionary mutation and crossover) with uncertainty-aware, sample-efficient selection (via DKL). The resulting waveforms enhance nonlinearity by repeatedly depinning the existing weakly pinned domain-wall segments of the as-grown ferroelectric, while avoiding the long-range macroscopic switching that would consume them. This constitutes a ferroelectric de-aging operation enacted through trajectory engineering rather than through bulk thermal or field treatment. In turn, it offers a route to in-situ recovery of the extrinsic piezoelectric response and to lower-voltage device operation. This framework provides a general route for autonomous design of driving protocols in scanning-probe microscopy and, more broadly, for machine-guided discovery of dynamic control strategies in functional materials.

Beyond this specific system, the results highlight a broader implication for process optimization: when the relevant design variable is a trajectory (a waveform, schedule, or recipe), the search space is inherently high dimensional, experimentally noisy, and strongly constrained by what can be executed. Methods that optimize only within conservative, "on-manifold" families can be effective for refinement, but they often miss the regimes where performance gains arise from a genuinely new protocol structure. By pairing an experimentally grounded prior with an evolutionary engine that can propose novel candidates and using uncertainty-aware learning to prioritize which candidates to test, the approach provides a practical blueprint for discovering processing histories that achieve the target functional outcome through repeated small-scale operations on the existing microstructure, rather than through a single large transformation that overwrites it. In the ferroelectric system studied here, this contrast is concretely realized by the depinning-dominated character of the high-ENL waveforms, which act on the as-grown weakly pinned wall population without driving the macroscopic switching events that would consume it.

More generally, this work supports a principle for real-world ML-driven discovery: discovery can be formulated as search in the space of experimental operations, not merely in the space of idealized structures or static descriptors. Real laboratories act through admissible actions such as finite sequences of controllable steps under constraints, and progress depends on algorithms whose proposal mechanisms and representations align with those actions. From this perspective, genetic operators defined over operational primitives offer a natural route to generate out-of-distribution candidates, while uncertainty-aware surrogates make that exploration tractable under measurement cost and noise. The resulting combination provides a pathway from optimization to out-of-distribution discovery that is directly executable in experimental workflows, and is broadly relevant to autonomous exploration across functional materials, processing routes, and other domains where the key unknown is not the object itself, but the program of operations that creates or controls it.

**Acknowledgements**

This work (development of GA-DKL, all the PFM experiments, and growth of the PZT sample) is supported by the center for 3D Ferroelectric Microelectronics Manufacturing (3DFeM$^2$), an Energy Frontier Research Center funded by the U.S. Department of Energy (DOE), Office of Science, Basic Energy Sciences under Award Number DE-SC0021118. C.-C.L. acknowledges that this material is based upon work supported by the Air Force Office of Scientific Research under award number FA9550-24-1-0266. Any opinions, findings, and conclusions or recommendations expressed in this material are those of the author(s) and do not necessarily reflect the views of the United States Air Force. J.K. acknowledges the support of the Army Research Office under the ETHOS MURI via cooperative agreement W911NF-21-2-0162. L.W.M. acknowledges the support of the Office of Naval Research under Grant ONR N000142612045.

## Methods

$Pb_{0.995}(Zr_{0.45}Ti_{0.55})_{0.99}Nb_{0.01}O_3$ films were grown by pulsed laser deposition using a KrF excimer laser from a ceramic target onto a $SrRuO_3$-electroded (001) SrTiO3 single crystal. The $SrRuO_3$ film was grown from a target from Kojundo Chemical Lab. Co. Ltd., using a laser energy density of 1.5 J/cm$^2$, a substrate temperature of 660°C, an oxygen pressure of 120 mTorr, a target-to-substrate distance of 6.7 mm, and a frequency of 5 Hz. The SrRuO3 film thickness was around 50 nm. The PZT film was grown from a target with 20% excess PbO to compensate for lead loss during growth, using a laser energy density of 1.5 J/cm$^2$, a substrate temperature of 630°C, an oxygen pressure of 120 mTorr, a target-to-substrate distance of 6.2 mm, and a frequency of 5 Hz. The PZT film thickness was around 147 nm.

Heterostructures PZTO-(111) sample consisting of 150 nm $PbZr_{0.2}Ti_{0.8}O_3$ and 30 nm $La_{0.67}Sr_{0.33}MnO_3$ (LSMO) were grown on $SrTiO_3$ (STO) substrates with (111) orientations (MTI Corp.) using pulsed-laser deposition (PLD). A KrF excimer laser (λ = 248 nm, LPX 300, Coherent) was employed to ablate ceramic targets with nominal compositions of $Pb_{1.2}Zr_{0.2}Ti_{0.8}O_3$ and $La_{0.67}Sr_{0.33}MnO_3$ (Praxair Inc.). Both the PZT and LSMO layers were deposited under identical conditions: a target-to-substrate distance of 60 mm, a substrate heater temperature of 635 °C, an oxygen partial pressure of 200 mTorr, a laser fluence of 1.5 J cm$^{-2}$, and a laser repetition rate of 4 Hz. After deposition, the samples were cooled to room temperature at a rate of 10 °C min$^{-1}$ in flowing oxygen at a pressure of ~700 Torr.

The SPM control is achieved by our home built open-source Python interface library, AESPM[47]. This library not only enables real-time operating the SPM system local or remotely with code the same way as human operators but also has access to the intermediate data like trace and retrace. AESPM is an open-source SPM-Python interface library. It can be found in the following link with detailed examples and tutorial notebooks: https://github.com/RichardLiuCoding/aespm

To help readers understand and reproduce the results in this work, we have provided an open-source Python notebook of applying GA-DKL on a real SPM instrument: https://github.com/RichardLiuCoding/Publications/tree/main/GA-DKL

**TOC Figure**

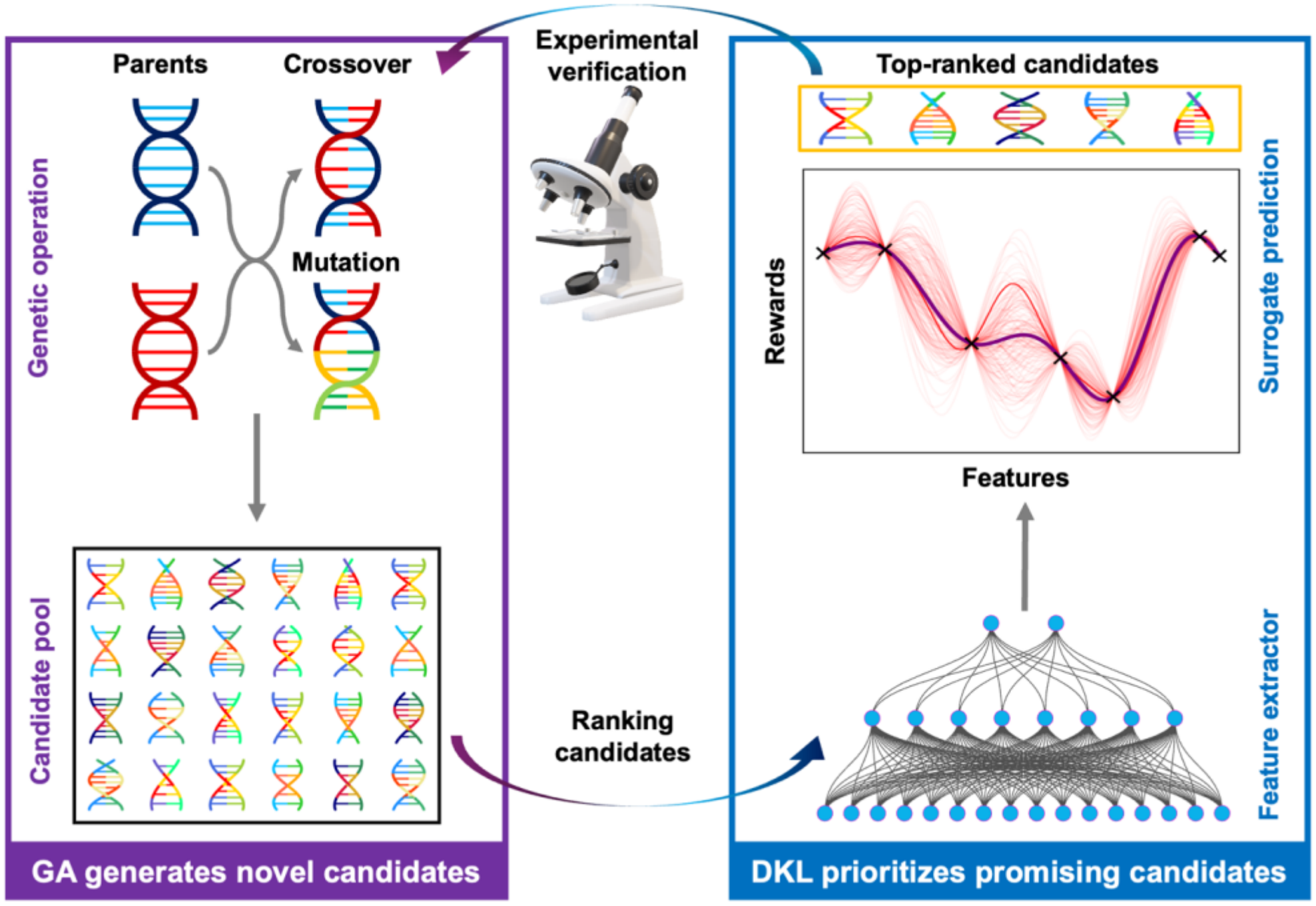

**TOC Figure.** Overview of the GA-DKL workflow. The genetic algorithm (GA) systematically generates diverse new candidate waveforms, including candidates outside the previously explored region, while deep kernel learning (DKL) trained on sparse experimental data predicts which candidates are most promising for measurement. Only the top-ranked candidates are tested experimentally, and the resulting data are fed back to improve the model in an iterative loop.

## Supplementary Materials

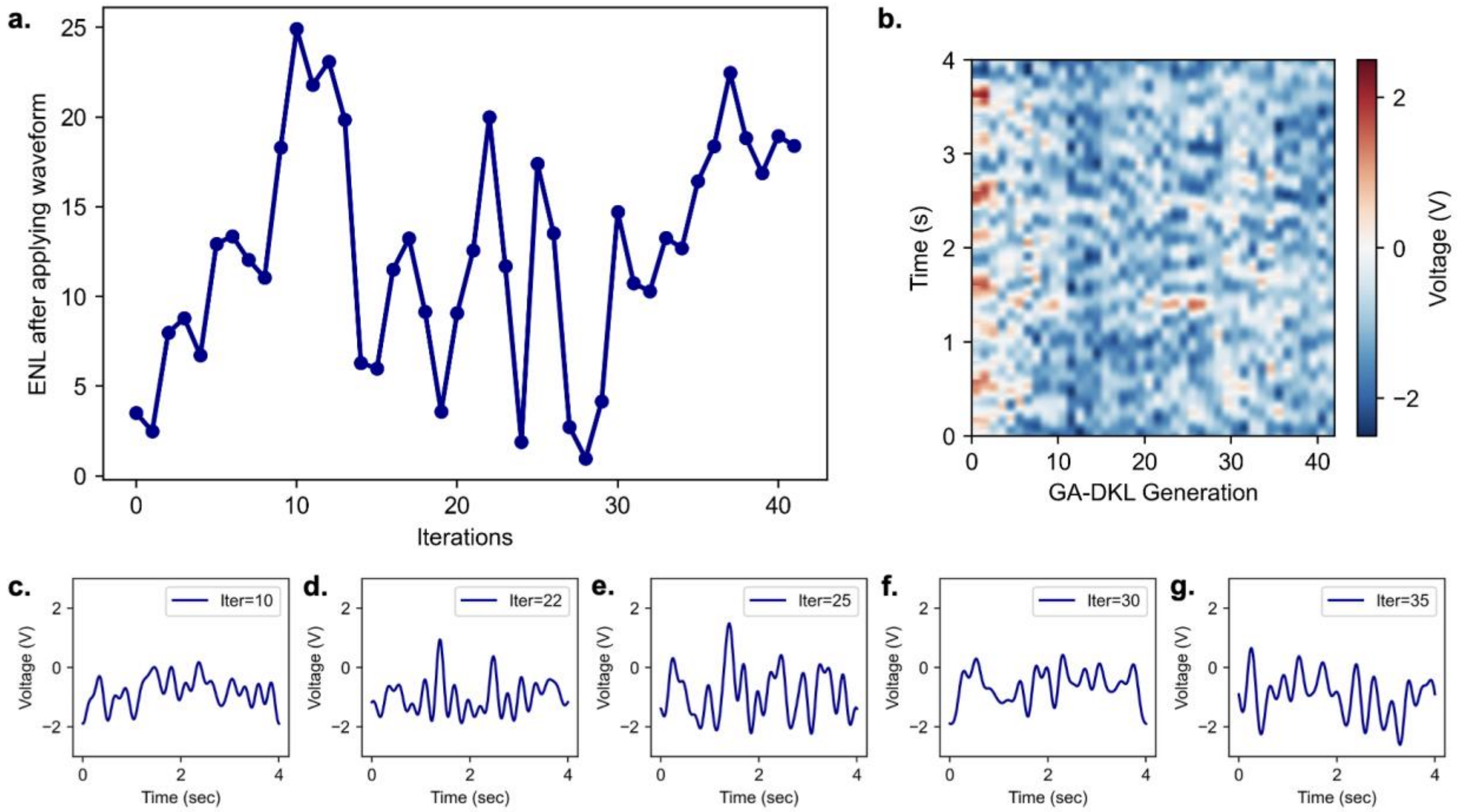

**Figure S1. Results of the GA-DKL experiment measured by the contact-tuning on PZT.** In this setup, the drive voltage was swept between 0.1 V and 1.2 V, and the contact tuning curve was obtained at each drive voltage. A Lorentzian fitting was used to extract the maximum piezoresponse at the resonance frequency in contact, and the ENL was extracted by the method shown in Fig. 1f.

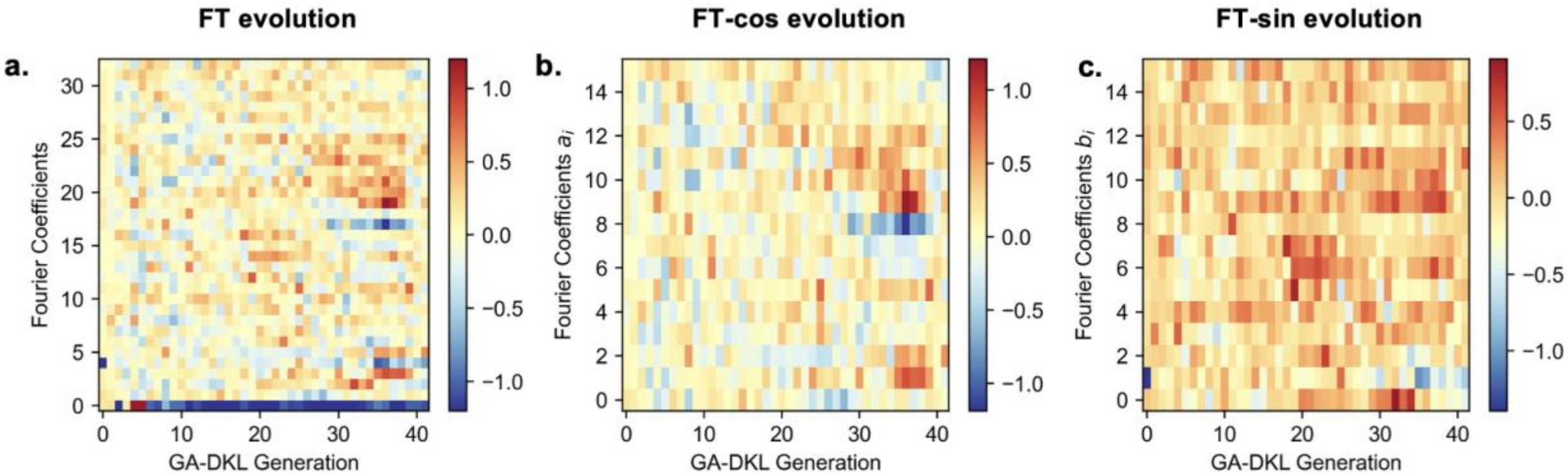

**Figure S2. Evolution of Fourier coefficients corresponding to the best waveforms.** Heat maps summarize how the Fourier representation of high-performing candidates changes across generations, highlighting the multi-harmonic character of the optimized waveforms.

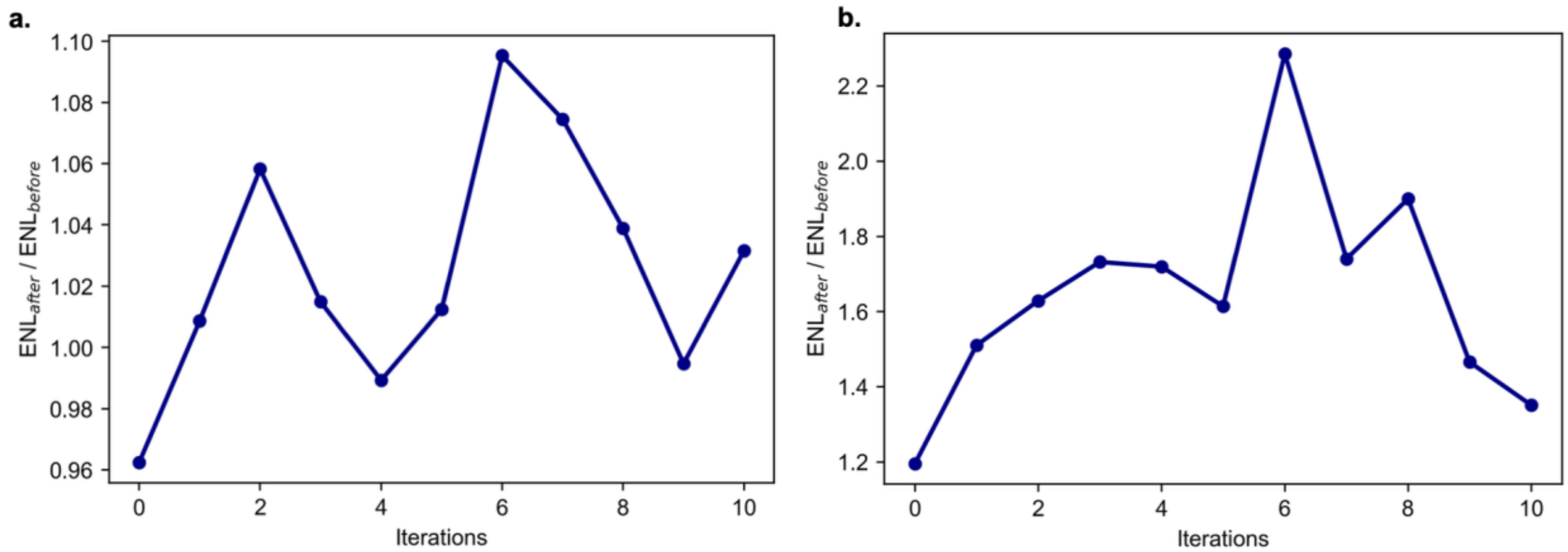

**Figure S3. Results of the GA-DKL experiment measured on $PbZr_{0.2}Ti_{0.8}O_3$ (PZTO-(111)) sample.** a, the averaged ENL ratio of each GA-DKL iteration. b, the evolution of the largest ENL ratio of each GA-DKL iteration. In this experiment, there are 2 seeding steps and 8 active learning iterations. Each step consists of 32 bias waveform candidates.

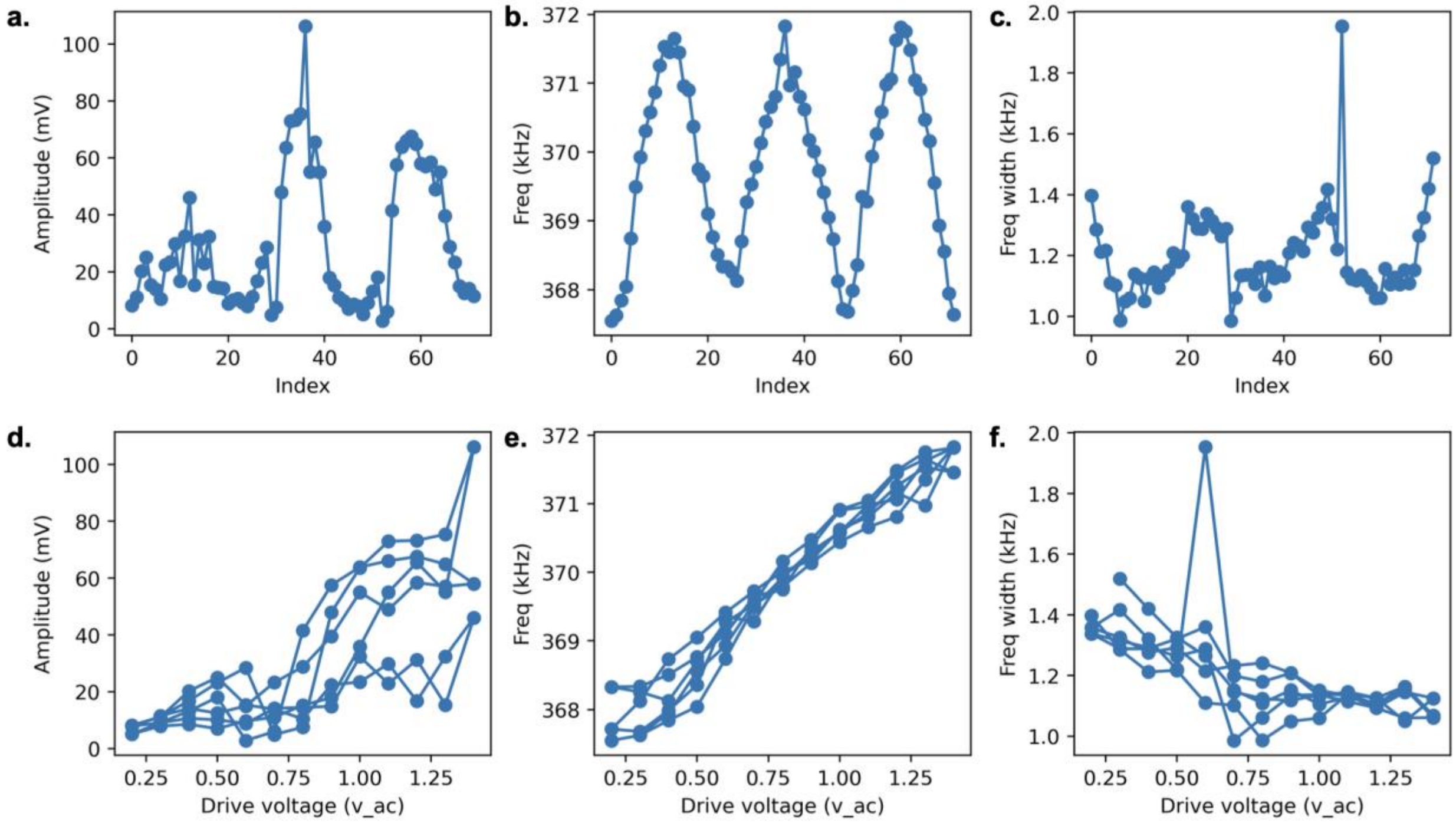

**Fig. S4. Reversible voltage-dependent shift of the contact resonance.** (a–c) Piezoresponse amplitude, contact-resonance frequency, and resonance peak width recorded sequentially during repeated drive-voltage sweeps in contact mode, plotted against measurement index. Each cycle ramps the AC drive voltage $V_{ac}$ up and back down, producing the periodic structure in (b). (d–f) The same data replotted against $V_{ac}$, with successive cycles overlaid. The contact-resonance

frequency shifts systematically by ~3–4 kHz between 0.2 and 1.4 V (b, e) and the resonance peak narrows modestly with drive (c, f), while the curves overlay closely from cycle to cycle, indicating that the shift is reversible and reflects a voltage-dependent change in the contact transfer function rather than irreversible sample modification.

**Algorithm 1. GA-DKL for active waveform optimization**

**Input:**

Initial measured archive $\mathcal{D}_0 = \{(x_i, y_i)\}$; experimental measurement function $\mathrm{MeasureENL}(x)$; batch size $B$; parent-pool size $P$; candidate-pool size $C$; number of generations $G$; exploration weight $\kappa$

**Output:**

Final archive $\mathcal{D}_G$; optimal waveform $V^*$

**Procedure:**

1. **Initialize the archive.**

Start from a seed set of experimentally measured waveforms and their corresponding nonlinear-response ratios:

$$y_i = \mathrm{MeasureENL}(x_i),$$
$$\mathcal{D}_0 = \{(x_i, y_i)\}.$$

2. **Represent waveforms in Fourier space.**

For each waveform $x(t)$, compute a truncated Fourier representation

$$x(t) = a_0 + \sum_{k=1}^{16} \left(a_k \cos\left(\frac{2\pi k}{T} t\right) + b_k \sin\left(\frac{2\pi k}{T} t\right)\right),$$

and define the coefficient vector

$$\theta = (a_0, a_1, b_1, \dots, a_{16}, b_{16}) \in \mathbb{R}^{33}.$$

3. **For each generation $g = 1, \dots, G$:**

a. Rank all measured waveforms in $\mathcal{D}_{g-1}$ by their measured response $y$.

b. Select the top $P$ waveforms as the parent set.

c. Apply GA crossover and mutation in Fourier-coefficient space to generate a candidate pool

$$\Theta_g = \{\theta_j\}_{j=1}^{C}.$$

d. Reconstruct the corresponding candidate waveforms

$$\mathcal{W}_g = \{x_j(t)\}_{j=1}^{C}.$$

e. Train a DKL surrogate on all previously measured pairs in $\mathcal{D}_{g-1}$ to predict the mean response $\mu_g(x)$ and predictive uncertainty $\sigma_g(x)$.

f. Score each candidate using the upper-confidence-bound acquisition function

$$A_g(x) = \mu_g(x) + \kappa\sigma_g(x).$$

g. Apply a diversity filter and select the top $B$ candidates.

h. Evaluate the selected waveforms experimentally:

$$y_j = \text{MeasureENL}(x_j).$$

i. Update the archive:

$$\mathcal{D}_g = \mathcal{D}_{g-1} \cup \{(x_j, y_j)\}.$$

4. **Return the optimum.**

The final optimal waveform is

$$x^* = \arg\max_{(x,y)\in\mathcal{D}_G} y$$